\documentclass{emulateapj}
\newcommand{\mr}{\mathrm}
\usepackage{graphicx,txfonts,natbib,verbatim}
\bibpunct{(}{)}{;}{a}{}{,}

\begin{document}

\title{Density fluctuations and the acceleration of electrons by beam-generated Langmuir waves in the solar corona}

\author{H. Ratcliffe}
\affil{School of Physics \& Astronomy, University of Glasgow, G12 8QQ, United Kingdom}
\email{h.ratcliffe@astro.gla.ac.uk}
\author{N. H. Bian}
\affil{School of Physics \& Astronomy, University of Glasgow, G12 8QQ, United Kingdom}
\author{E. P. Kontar}
\affil{School of Physics \& Astronomy, University of Glasgow, G12 8QQ, United Kingdom}

\begin{abstract}
Non-thermal electron populations are observed throughout the
heliosphere. The relaxation of an electron beam is known to produce
Langmuir waves which, in turn, may substantially modify the electron
distribution function. As the Langmuir waves are refracted by
background density gradients and as the solar and heliospheric
plasma density is naturally perturbed with various levels of
inhomogeneity, the interaction of Langmuir waves with non-thermal
electrons in inhomogeneous plasmas is an important topic. We
investigate the role played by ambient density fluctuations on the
beam-plasma relaxation, focusing on the effect of acceleration of
beam electrons. The scattering of Langmuir waves off turbulent
density fluctuations is modeled as a wavenumber diffusion process
which is implemented in numerical simulations of the one-dimensional
quasilinear kinetic equations describing the beam relaxation. The
results show that a substantial number of beam electrons are
accelerated when the diffusive time scale in wavenumber space
$\tau_{D}$ is of the order of the quasilinear time scale
$\tau_{ql}$, while when $\tau_{D}\ll \tau_{ql}$, the beam relaxation
is suppressed. Plasma inhomogeneities are therefore an important means of
energy redistribution for waves and hence electrons, and so must
be taken into account when interpreting, for example, hard X-ray or
Type III emission from flare-accelerated electrons.
\end{abstract}




\maketitle
\section{Introduction}

The relaxation of an electron beam producing Langmuir waves is a
fundamental process in space and laboratory plasmas. As the Langmuir
waves interact with the fast electrons, and their dynamics are
strongly affected by variations in the ambient plasma density and as
the heliospheric plasma density is perturbed with different levels
of inhomogeneity, the interaction of Langmuir waves with non-thermal
electrons in turbulent inhomogeneous plasmas is an important topic
in astrophysics.

In the solar atmosphere, the beam-plasma instability is known to be
the cause of Type III radio bursts, with their characteristic rapid
frequency drift being a result of the beam propagation through the
decreasing density coronal plasma
\citep{1958SvA.....2..653G,1970SvA....14...47Z,1990SoPh..130....3M}. 
Numerical simulations of the beam relaxation in the inhomogeneous plasma of
the solar wind can reproduce the broken power-law in the electron
flux spectrum which is observed near the Earth for impulsive solar
electron events \citep{2010ApJ...721..864R}. In solar flare
environments, it has been shown that the beam relaxation is likely
to cause spectral index variations inferred from hard X-ray
observations \citep[e.g.][]{2009ApJ...707L..45H}.

It was recognized already in the context of beam-plasma experiments
\citep{1969JETP...30..759B, 1965JAMTP...6....9L} that the mere
existence of an inhomogeneity in the ambient plasma density was
sufficient to produce a redistribution of Langmuir wave energy in
wavenumber space. This idea was proposed in order to explain the
common observations of electrons with energy above the injected one
\citep{1969JETP...30..759B} in laboratory experiments such as those
carried out by \citet{1964JNuE....6..173B} on powerful relativistic
beams. The interaction between the waves and the beam electrons is
resonant, which implies that the spectral transfer of wave energy
toward lower/higher wavenumbers allows interaction with
faster/slower electrons respectively. Various mechanisms may cause
this spectral energy transfer, they may involve regular or random
density variations, or non-linear interactions between Langmuir
waves and low-frequency electrostatic compressive modes. For
example, it is known that ion-sound waves can effectively scatter
Langmuir waves to smaller wavenumbers
\citep[e.g.,][]{1965JAMTP...6....9L,1974SoPh...35..441M,1975PhRvL..35..995E,1975PhFl...18.1769P,1978A&A....68..405E}.
Low-frequency electromagnetic modes, such as kinetic Alfven waves
\citep{2010PhPl...17f2308B,2010A&A...519A.114B} are compressive and
thus their refractive power may also have a substantial effect on
the spectral evolution of beam-driven Langmuir waves. Numerical
simulations of the beam relaxation in a plasma with a constant
density gradient were conducted in
\citep[e.g.][]{1978SvJPP...4R1267K,2001A&A...375..629K,2010SoPh..267..393T}
and the combination of a regular density gradient with allowance for
the non-linear coupling to ion sound waves was considered by
\citet{2002PhRvE..65f6408K,2006JGRA..11109106Y,2011ApJ...727...16Z},
with the result that a very high level of Langmuir waves was
produced at small wavenumbers. The simulations by
\citet{2012A&A...539A..43K} found that a fluctuating density could
lead to the appearance of accelerated electrons in a
collisionally-relaxing electron beam.

\citet{1967PlPh....9..719V} first described the effect of random
large scale density inhomogeneities as a diffusive transfer of
Langmuir wave energy in wavenumber space. This diffusion equation
was used by \citet{1976JPSJ...41.1757NF} to describe the process of
elastic scattering off random and time-independent density
fluctuations, \citep[see
also][]{1982PhFl...25.1062G,1991PhFlB...3.1968M}. Elastic scattering
results only in angular diffusion in wave-number space, and for the
case of waves generated by beam electrons with a small angular
spread, wave energy is transferred away from the region of
excitation in $k$-space, which may lead to suppression of the beam-plasma instability. Neglecting the effect of wave reabsorption on the electrons, the role of the angular diffusion term has been accounted for in the kinetic equations describing the beam-plasma system, by replacing it by a wave damping term \citep{1985SoPh...96..181M,1987SoPh..111...89M}. However, in general, energy conservation dictates that reabsorption of wave energy by the beam is accompanied by an acceleration of the electrons. In addition, inelastic scattering results in a change in the absolute value of the Langmuir wavenumber, which may produce acceleration in the projected electron distribution.

Here, we consider anew this important topic of beam relaxation in a
fluctuating plasma. In Section II, we derive a general expression
for the diffusion coefficient in wavenumber space. In Section III,
the wavenumber diffusion is implemented in numerical simulations of
the one-dimensional kinetic equations describing the quasilinear
evolution of the beam-plasma instability. We focus on the effect of
acceleration of beam electrons in a fluctuating plasma and derive a
condition for it to occur. In Section IV, we generalize the
discussion to 3D and discuss the role of both elastic and inelastic
scattering in the acceleration of fast electrons. A summary of the
results is presented in Section V.


\section{Langmuir wave scattering and diffusion: 1D case}
Let us start by considering a one-dimensional problem where the dynamics of
both electrons and Langmuir waves are along $x$, the direction of the external
magnetic field. The plasma density is written as $n_e[1+\tilde{n}(x, t)]$ with $n_e$
the constant background density and $\tilde{n}(x,t)$ the relative density fluctuation,
which is assumed to be weak, i.e. $\tilde{n}(x,t)\ll1$, a condition often satisfied in
the solar corona and the solar wind.

In general, the characteristic wavenumber $q$ of low-frequency
density fluctuations is much smaller than the characteristic
wavenumber $k$ associated with high-frequency Langmuir waves, so we
can make the WKB approximation and treat the Langmuir waves as
quasi-particles. Ambient density fluctuations are associated with a
change in the local refractive index experienced by the waves, and
in the low frequency limit the Langmuir wave dynamics can thus be
modeled by Hamilton's equations of motion for the 
quasi-particles \citep[e.g.][]{1965JFM....22..273W,1967PlPh....9..719V,1974R&QE...17..326Z},
which are given by
\begin{equation}\label{eqn:WKB1} \frac{d k}{d t} =-\frac{1}{2}
\omega_{pe} \frac{\partial\tilde{n}}{\partial x}\equiv F(x,t)
\end{equation}
\begin{equation}
\frac {d x}{d t} ={\mr v}_g
\end{equation}
where $F(x,t)$ is the "refraction force" acting on the wave-packets,
${\mr v}_g=3{\mr v}_{Te}^2 k/\omega_{pe}$ is the group velocity and
$\omega_{pe}$ is the local plasma frequency $\omega_{pe}=\sqrt{4\pi
n_e e^2/m_e}$.

Equivalently, we can write a conservation relation describing the
evolution of the spectral energy density associated with Langmuir
waves $W(x, k, t)$ [erg cm$^{-2}$]
\begin{equation}
\label{eqn:liou}
\frac {\partial W(x, k, t)}{\partial t} +{\mr v}_g\frac{\partial W(x, k, t)}{\partial x}
-
F(x,t)\frac{\partial W(x, k, t)}{\partial k}=0,
\end{equation}
where $W(x,k,t)$ is normalised to the energy density of
Langmuir waves $E=\int W(x, k, t)dk$.

From Equation (\ref{eqn:WKB1}) we see that random refraction induces
stochastic change in the wavenumber, resulting in diffusion of the
spectral energy density in $k$-space. An expression for the
diffusion coefficient $D$ is derived from the phase-space
conservation equation (\ref{eqn:liou}) following standard procedures
\citep[e.g.][]{1963JETP...16..682V,1966PhRv..141..186S}, as follows.

The spectral energy density of Langmuir waves $W(x,k,t)$
is decomposed into the sum of its average and fluctuating parts as $ W=\langle W\rangle  +\widetilde{W}$.
Substituting this expression into Eq. (\ref{eqn:liou})
gives one equation for the average
\begin{equation}\label{eqn:mean}
\frac{\partial \langle W\rangle}{\partial t}
=-\langle F(x,t)\frac{\partial \widetilde{W}}{\partial k}\rangle
\end{equation}
and one equation for the fluctuations
\begin{equation}\label{eqn:fluct}
\frac {\partial \widetilde{W}}{\partial t}+{\mr v}_g\frac{\partial \widetilde{W}}{\partial x}=-F(x,t)\frac{\partial \langle W \rangle}{\partial k}
\end{equation}
where a term quadratic in the fluctuation amplitude
has been neglected since the latter is assumed to be weak.
Equation (\ref{eqn:fluct}) is integrated to give
\begin{equation}\label{eqn:pert_sol}
\widetilde{W}(x,k,t)=-\int_{0}^t d\tau F(x-{\mr v}_g\tau, t-\tau)\frac{\partial \langle W\rangle(x-{\mr v}_g\tau, k, t-\tau)}{\partial k},
\end{equation} which is substituted into Eq. (\ref{eqn:mean}) to give the equation describing the diffusion of wave energy in $k$-space,
\begin{equation}
\frac{\partial \langle W\rangle}{\partial t}
=\frac{\partial }{\partial k}D\frac{\partial \langle W\rangle}{\partial k},
\end{equation}
where the diffusion coefficient $D$ is expressed in terms of the
auto-correlation function of the refraction force as
\begin{equation}\label{eqn:diff2}
D=\int_0^\infty d\tau \langle F(x, t)F(x-{\mr v}_g\tau, t-\tau)\rangle.
\end{equation}
The Fourier components $F(q, \Omega)$ and the spectrum $S_F(q,\Omega)$ of $F(x, t)$ are
defined through $ F(x, t)=\int_{-\infty}^\infty dq\int_{-\infty}^\infty d\Omega F(q, \Omega) \exp{[2\pi i (qx -\Omega t)]} dk d\Omega$ and $\langle F(q,\Omega) F(q', \Omega')\rangle= S_F(q, \Omega)\delta (q+q') \delta (\Omega+\Omega')$, respectively. As a consequence, the diffusion coefficient can be written in terms of $S_F(q,\Omega)$ as \begin{equation}\label{Dv2}
 D=\frac{1}{2}\int_{-\infty}^\infty dq\int_{-\infty}^\infty d\Omega S_F(q, \Omega)\delta(\Omega-q{\mr v}_g).
 \end{equation}
Since the spectrum of the refraction force $S_F(q,\Omega)$ is
related to the spectrum of ambient plasma density fluctuations
$S_n(q,\Omega)$ by $S_F(q, \Omega)=\omega_{pe}^2(\pi q)^2 S_n(q,
\Omega)$, the diffusion coefficient can finally be expressed as
\begin{equation}\label{eqn:diff_1d} D= \frac{\omega_{pe}^2 \pi^2}{2}
\int_{-\infty}^\infty dq \int_{-\infty}^\infty d\Omega\; q^2 S_n(q,
\Omega)\delta(\Omega-q{\mr v}_g),
\end{equation}
where by definition
$\langle\tilde{n}^2\rangle=  \int_{-\infty}^\infty dq \int_{-\infty}^\infty d\Omega\; S_n(q,\Omega)$.

According to Eq. (\ref{eqn:diff_1d}), a quasi-particle with
wave-number $k$ interacts with a Fourier mode $(\Omega,q)$ of the
density fluctuation spectrum for which the resonance condition
$\Omega=q {\mr v}_g$ is satisfied, which also means that
$(\Omega/q)(\omega_{pe}/k) = 3 {\mr v}_{Te}^2$. In the particular
case where density fluctuations are due to waves with a dispersion
relation $\Omega=\Omega(q)$, then
$S_n(q,\Omega)=S_n(q)\delta(\Omega-\Omega(q))$ and $ D=
(\omega_{pe}^2 \pi^2/2)\int_{-\infty}^\infty dq q^2
S_n(q)\delta(\Omega(q)-q{\mr v}_g)$.

For a Gaussian spectrum
\begin{equation}  S_n(q, \Omega)=\frac{\langle\tilde{n}^2\rangle }{\pi q_0\Omega_0} \exp\left(-\frac{q^2}{q_0^2}-\frac{\Omega^2}{\Omega_0^2}\right)
\end{equation} where $q_0$ and $\Omega _0$  are characteristic wavenumber and frequency, the diffusion coefficient
reads
\begin{equation}\label{eqn:diffcoeffGauss}
D= \omega_{pe}^2{\pi}^{3/2}\frac{q_0}{{\mr v}_0}\langle\tilde{n}^2\rangle \left(1+\frac{{\mr v}_g^2}{{\mr v}_0^2}\right)^{-3/2},
\end{equation}
where ${\mr v}_0=\Omega_0/q_0$.

The diffusion coefficient depends on wave-number $k$ through the
group velocity ${\mr v}_{g}=3{\mr v}_{Te}^{2}k/\omega_{pe}$. From
Eq. (\ref{eqn:diffcoeffGauss}), we see that for a Gaussian spectrum
there are two regimes of wavenumber diffusion. When the
decorrelation velocity associated with ambient density fluctuations
${\mr v}_{0}$ is much larger than the group velocity of the Langmuir
waves, i.e. when ${\mr v}_{0}\gg {\mr v}_{g}$, then
\begin{equation} \label{eqn:Dlim1} D= \omega_{pe}^2{\pi}^{3/2}
\left(\frac{q_{0}}{{\mr v}_{0}}\right)\langle\tilde{n}^2\rangle,
\end{equation}
an expression which is independent of ${\mr v}_{g}$ and hence $k$.
Conversely, when ${\mr v}_{0}\ll {\mr v}_{g}$, then
\begin{equation}\label{eqn:Dlim2}
D= \omega_{pe}^2{\pi}^{3/2} \left(\frac{q_0}{{\mr v}_{0}}\right)
\left(\frac{{\mr v}_{0}}{{\mr v}_{g}}\right)^{3}\langle\tilde{n}^2\rangle,
\end{equation}
and the wavenumber diffusion becomes $k$ dependent. 

It is also
interesting to consider compressive fluctuations at a given
characteristic frequency, so that
$S_n(q,\Omega)=S_n(q)\delta(\Omega-\Omega_0)$, with a power-law
spectrum in wavenumber space with spectral index $\zeta
>1$, as observed in the solar wind
\citep[e.g.][]{1972ApJ...171L.101C,1983A&A...126..293C,1983PASAu...5..208R}. Then
\begin{equation} S_n(q,\Omega)=
\langle\tilde{n}^2\rangle\left(\frac{\zeta-1}{q_0}\right)\left(\frac{q}{q_0}\right)^{-\zeta}\delta(\Omega-\Omega_0)\end{equation}
for $q >q_0$ and zero elsewhere. In this case
we find that
\begin{equation}
D=\frac{\omega_{pe}^2\pi^2}{2}(\zeta-1)\left(\frac{q_0}{{\mr v}_0}\right)
\left(\frac{{\mr v}_0}{{\mr v}_g}\right)^{3-\zeta}\langle\tilde{n}^2\rangle.
\end{equation}
Comparison of this last expression with Eq. (\ref{eqn:Dlim2}) shows
how the spectrum of density fluctuations affects the wavenumber
dependence of the diffusion coefficient $D$.

\section{Numerical results}

We now perform numerical simulations of the quasilinear kinetic equations describing wave-particle interactions.
These equations are based on those given by \citet{1963JETP...16..682V,1967PlPh....9..719V,1995lnlp.book.....T}
adding a spontaneous wave emission term as in \citet{1970SvA....14...47Z,2009ApJ...707L..45H,2011A&A...529A..66R}
and collisional operators as in e.g \citet{1981phki.book.....L}:
\begin{equation}\label{eqn:ql1}
\frac{\partial f}{\partial t}= \frac{4\pi^2
e^2}{m_e^2}\frac{\partial}{\partial {\mr v}}\left( \frac{W}{{\mr
v}}\frac{\partial f}{\partial {\mr v}}\right) +
\Gamma\frac{\partial}{\partial \mr v}\left(\frac{f}{\mr
v^2}+\frac{\mr v_{Te}^2}{\mr v^3}\frac{\partial f}{\partial \mr
v}\right),\;\;\; \label{eqk1}
\end{equation}
\begin{equation}\label{eqn:ql2}
\frac{\partial W}{\partial t} =\frac{\omega_{pe}^3 m_e}{4\pi n_e}\mr
v \ln\left(\frac{\mr v}{\mr v_{Te}}\right)
+\frac{\pi\omega_{pe}^3}{n_ek^2}W\frac{\partial f}{\partial {\mr
v}}-\frac{\Gamma}{4 {\mr v}_{Te}^3}W+\frac{\partial}{\partial
k}\left(D\frac{\partial W}{\partial k}\right), \label{eqk2}
\end{equation}
where $\Gamma=4 \pi e^4 n_e \ln \Lambda/m_e^2$, with
$\ln\Lambda\simeq 20$ the Coulomb logarithm.
The last term in Eq. (\ref{eqn:ql2}) describes the effect of wavenumber diffusion produced by the
ambient density fluctuations, as discussed in the previous section.

The initial electron distribution function $f({\mr v},t)$ [electrons
cm$^{-3}$ (cm/s)$^{-1}$ ] is chosen as the superposition of a
Maxwellian background and a Maxwellian beam:
\begin{equation}
f({\mr v}, t=0)= \frac{n_e}{\sqrt{2\pi} {\mr v}_{Te}}
\exp\left(-\frac{{\mr v}^2}{2 {\mr v}_{Te}^2}\right) +\frac{n_{b} }{\sqrt{\pi} \Delta {\mr v}_b} \exp\left(-\frac{({\mr v}-{\mr v}_b)^2}{\Delta {\mr v}_b^2}\right)
\end{equation}
and the initial spectral energy density of Langmuir waves $W(k,t)$ [ergs
cm$^{-2}$ ] is set to the thermal level:
\begin{equation}
W(k, t=0)= \frac{k_b T_e}{4 \pi^2} k^2\ln\left(\frac{1}{k\lambda_{de}}\right).
\end{equation}

Except where otherwise stated, the simulations below assume a
background plasma with $T_e=T_i=1$MK and $\omega_{pe}/2\pi=1GHz$
giving a background density of $n_e=1.2\times 10^{10}$cm$^{-3}$. The
beam parameters are $n_b=10^5\simeq 10^{-5} n_e$, ${\mr
v}_b=5\times10^9$cms$^{-1}$ and $\Delta{\mr v}_b=0.3{\mr v}_b$. The
spectrum of background density fluctuations is taken first to be
Gaussian, so the diffusion coefficient $D$ is given by
Eq. (\ref{eqn:diffcoeffGauss}). We fix $q_0=10^{-4} k_{De}$. The
magnitude of relative density fluctuations $\sqrt{\langle
\tilde{n}^2\rangle}$ ranges between $10^{-4}$ and $10^{-1}$, and
${\mr v}_0$ is taken between $0.01{\mr v}_{Te}$ and ${\mr v}_{Te}$.

\begin{figure}
\center
\includegraphics[width=0.45\linewidth]{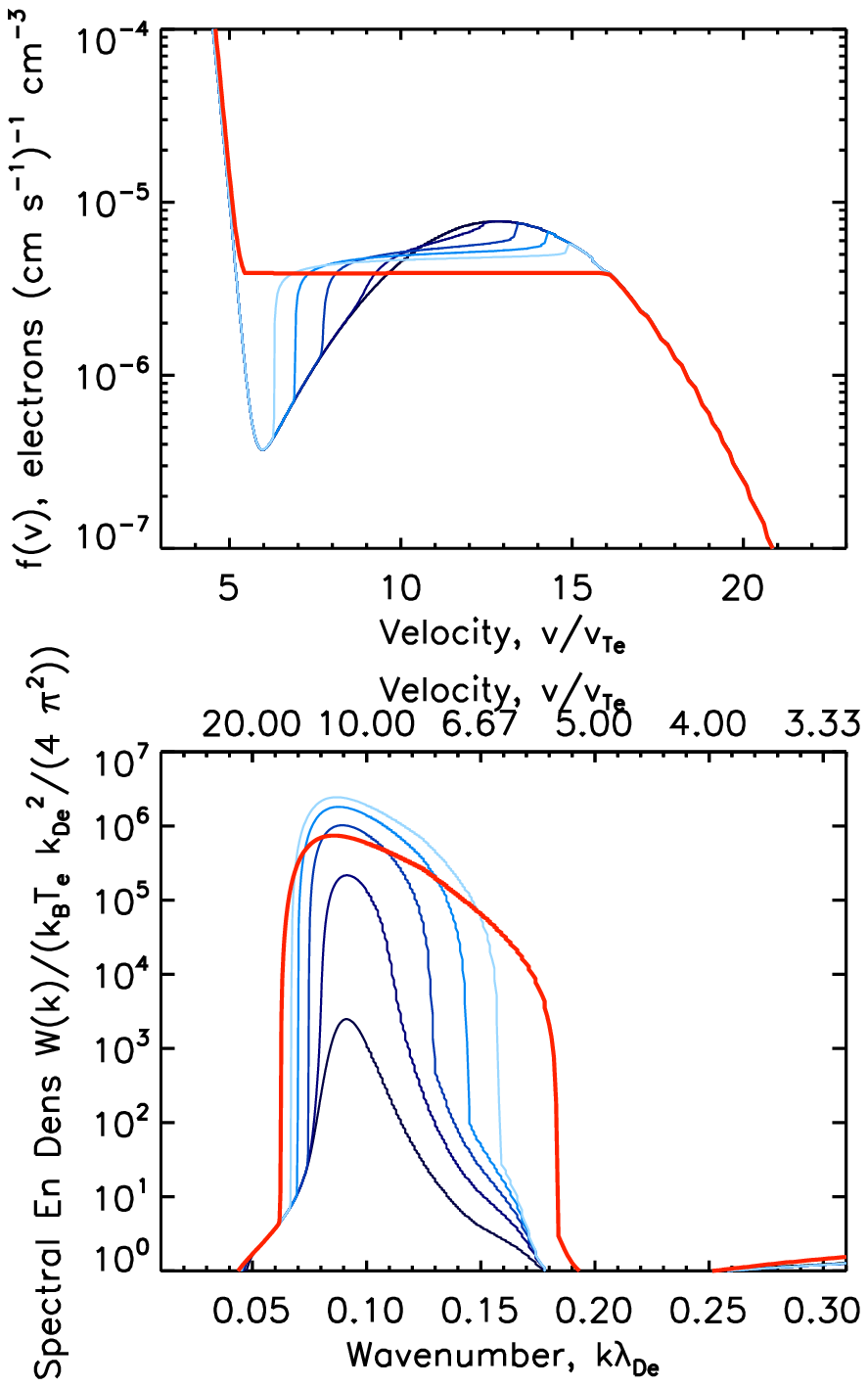}
\includegraphics[width=0.45\linewidth]{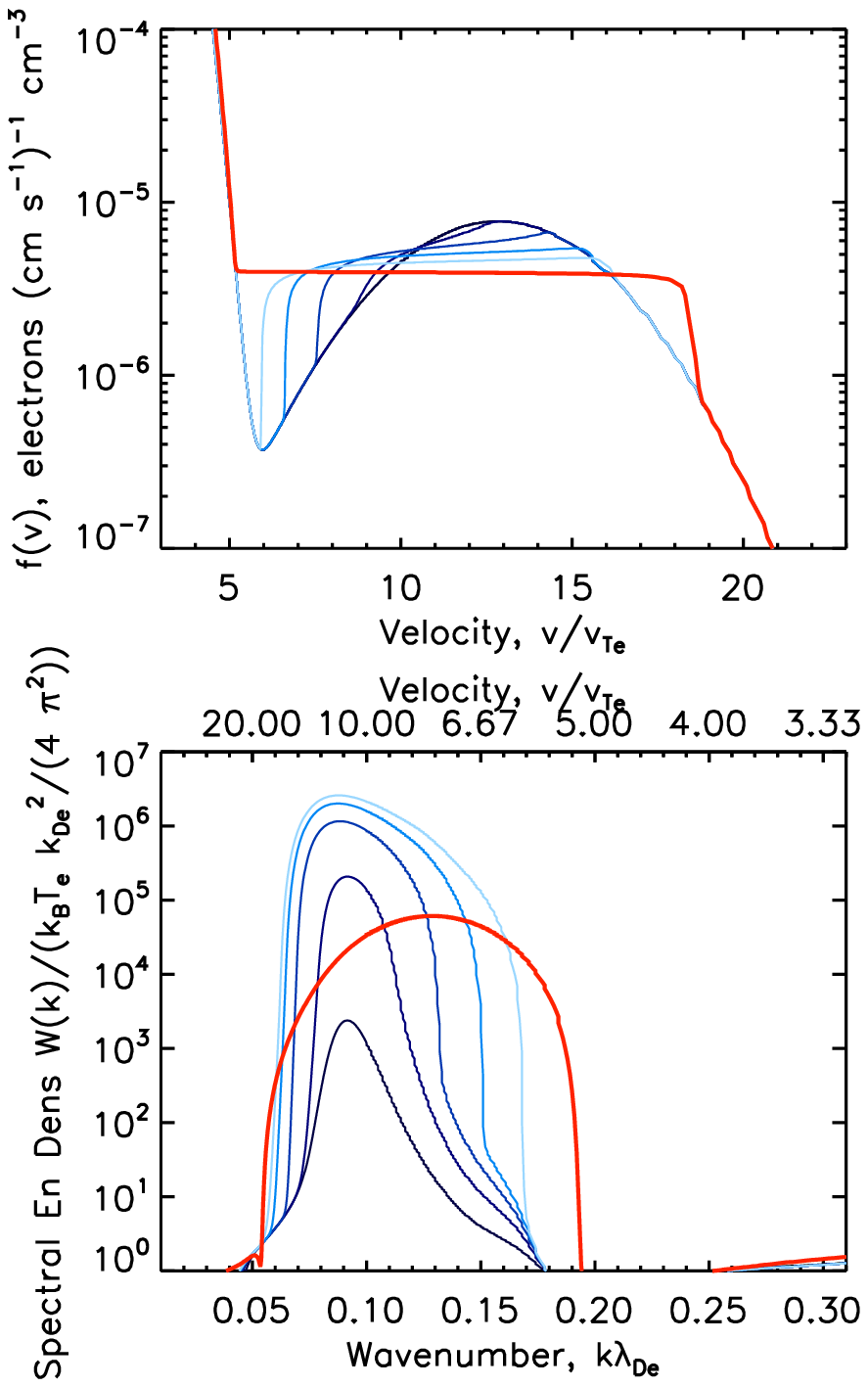}
\includegraphics[width=0.45\linewidth]{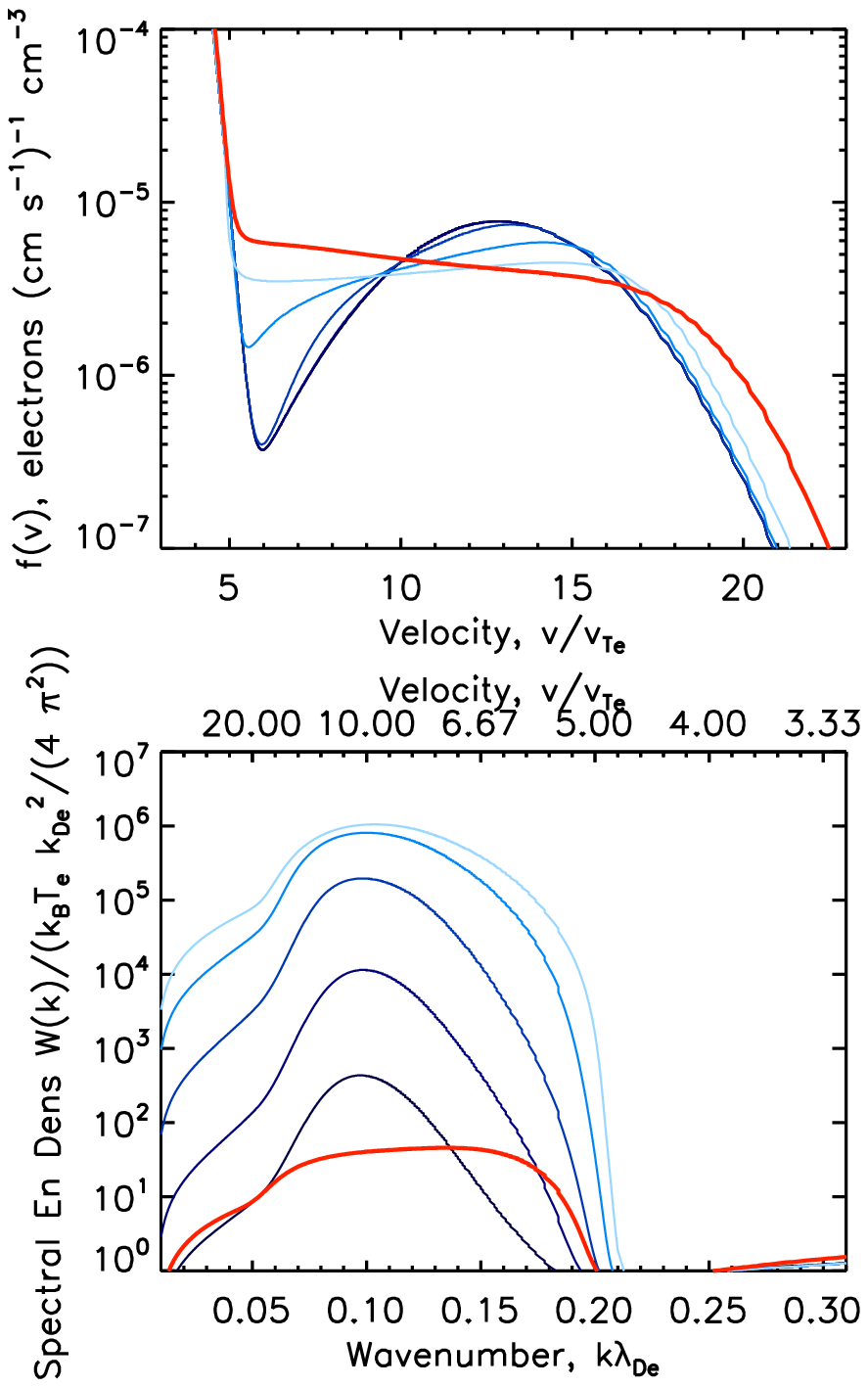}
\includegraphics[width=0.45\linewidth]{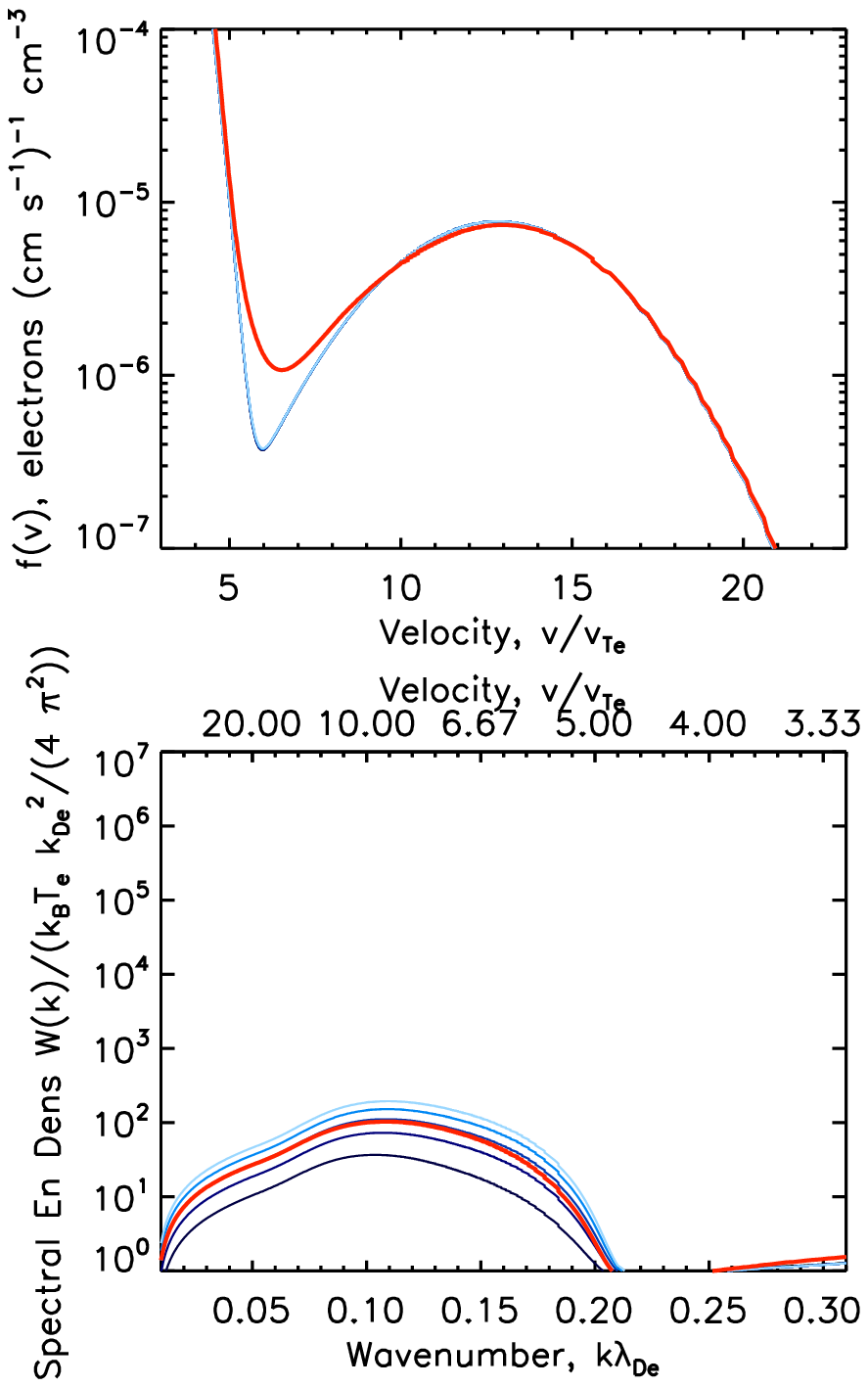}
\caption{Electron distribution $f(\mr v)$ (upper plots), and Langmuir wave spectral energy density $W(k)$ (lower plots) for homogeneous (top left) and inhomogeneous plasma, with $q_0=10^{-4}k_{de}$, ${\mr v}_0/{\mr v}_{Te}=0.3$ and $\sqrt{\langle \tilde{n}^2\rangle}$ of $3.7\times10^{-4}$ (top right), $3.7\times10^{-3}$ (bottom left) and $1.2\times10^{-2}$ (bottom right). Lines from dark (blue in online version) to light show the beam relaxation during the first ten quasilinear times, while the thick line (red in online version) shows the state reached at $t=100\tau_{ql}$.}\label{fig:w_wo}
\end{figure}

Figure (\ref{fig:w_wo}) is divided into four distinct panels which
all show the evolution of the electron and the wave distributions as
the intensity of the background density fluctuations $\sqrt{\langle
\tilde{n}^2\rangle}$ is increased. In a homogeneous plasma, as shown
in the top left panel, the electron beam is unstable to the
generation of Langmuir waves, which grow at a rate given by
\begin{equation}
\gamma\sim \frac{n_b}{n_e}\omega_{pe}.
\end{equation}
The spectrum of Langmuir waves interacts with the beam electrons and
causes them to diffuse in the resonant region in velocity space, given
by $\omega_{pe}=k{\mr v}$, until the distribution function reaches a
plateau, on a time scale given by the quasilinear time:
\begin{equation}
\tau_{ql}=\frac{1}{\omega_{pe}}\frac{n_{e}}{n_{b}}.
\end{equation}
By time $t=100 \tau_{ql}$, the beam has fully relaxed and the
electron distribution is flat between $6 {\mr v}_{Te}$ ($E =
1.5$~keV) and $16 {\mr v}_{Te}$ ($11$~keV), as shown by the red lines in Figure \ref{fig:w_wo}.
The collisional timescale, $\tau_{coll}={\mr v}_{Te}^3/\Gamma$ and
the timescale for collisional destruction of the plateau,
$\tau_{coll}({\mr v/{\mr v}_{Te}})^3\sim 1$~s at ${\mr v}=15 {\mr v}_{Te}$ 
are both far longer than the quasilinear time. Collisional effects are
therefore negligible. This remains true for the various values of
beam density and background plasma parameters which will be
considered below.

A finite level of background density fluctuations produces diffusion
of the waves in $k$-space which allows them to interact resonantly
with electrons from a larger region of velocity space. In other
words, the region of resonant wave-particle interaction is broadened
due to the random wave refraction induced by the ambient density
perturbations. This effect is well illustrated by the bottom left
panel of Figure \ref{fig:w_wo} which shows the electron and wave
distributions for a moderate level of density fluctuations,
$\sqrt{\langle \tilde{n}^2\rangle}=3.7\times 10^{-3}$. The Langmuir
wave spectrum diffuses in wavenumber, widening the beam distribution
and slowing the plateau formation. The electron distribution
$f({\mr v})$ is increased from the upper edge of the plateau to the
highest velocities in the simulation, implying that electrons have
been accelerated. In addition, the increased level of waves at large
wavenumbers has a visible effect on the electron distribution down
to around $6{\mr v}_{Te}$.

The top right panel shows very weak density fluctuations, with
$\sqrt{\langle \tilde{n}^2\rangle}=3.7\times 10^{-4}$. The beam
relaxes in a manner similar to the homogeneous case but the wave
spectrum is broadened slightly in $k$-space, resulting in a wider
plateau in the electron distribution.

When the inhomogeneity becomes very strong, as illustrated by the
bottom right panel where $\sqrt{\langle
\tilde{n}^2\rangle}=1.2\times 10^{-2}$, diffusion still transports
the waves out of their region of excitation in $k$-space but on a
timescale which is much smaller than their growth rate, so the wave
level is barely increased above the thermal level. As we can see, by time $t=100\tau_{ql}$ the electron distribution remains essentially unchanged from the initial distribution. The main point
is that for intense density fluctuations, diffusive broadening of
the Langmuir wave spectrum becomes large enough for the waves to be
mainly reabsorbed by the thermal electrons at $k\lambda_{De}\sim
1$. Since the energy density of waves generated is much
smaller than the thermal energy of plasma, the effect on the
distribution function of thermal particles is negligible. 
This is the suppression regime of the beam-plasma instability which has
often been discussed in the past, mainly in the context of the role
of angular diffusion of wave energy induced by elastic scattering in
3D. As shown here, the suppression of the beam-plasma instability can
also occur as a result of inelastic scattering. Here, we focus on
the effect of electron acceleration due to wave reabsorption by the
beam electrons as opposed to reabsorption by the the thermal
electrons. A discussion of electron acceleration in 3D, as a result
of both elastic and inelastic scattering, is presented in the next
section.

Clearly, the main effect of increasing the level of density
fluctuations is to increase the rate at which wave energy is
transported in $k$-space. The time scale
 associated with such diffusive process is 
 \begin{equation}
 \tau_D = \frac{(\Delta k_{b})^2}{D(k_b)}
\end{equation}
where $k_{b}=\omega_{pe}/{\mr v}_{b}$ and $\Delta k_{b}=\omega_{pe} (\Delta {\mr v}_{b}/{\mr v}_{b}^{2})$ , 
with parameters $\omega_{pe}, {\mr v}_b$ and $\Delta {\mr v}_b$ as given above. Therefore, it is natural to define a non-dimensional number
$R$, which is the ratio of the diffusion time $\tau_{D}$ and the quasilinear time $\tau_{ql}$, i.e.
\begin{equation}
R=\frac{\tau_{D}}{\tau_{ql}}.
\end{equation}

\begin{figure}

\includegraphics[width=0.98\linewidth]{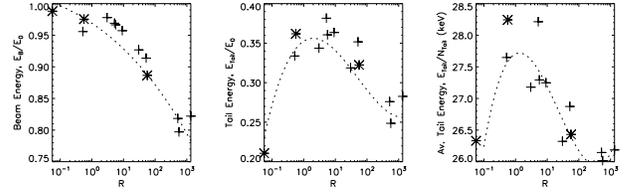}
  \caption{The total energy of beam (left) and tail (middle) electrons, both normalised to the initial beam energy $E_0$, and the average energy of a tail electron (right), all evaluated at $t=100\tau_{ql}$, against the parameter $R=\tau_D/\tau_{ql}$, which gives a measure of the degree of inhomogeneity. Beam and plasma parameters, and ranges for $\sqrt{\langle
\tilde{n}^2\rangle}$ and ${\mr v}_0$ are as given in the text. The asterisks mark the three levels of inhomogeneity shown in Figure \ref{fig:w_wo}. The dashed lines are polynomial fits to the data to illustrate its general trends. }\label{fig:parst}
\end{figure}

To quantify things further, let us define a ``beam'' region in
velocity space that includes all electrons with velocity above $6
{\mr v}_{Te}$, and a ``tail'' region above $16{\mr v}_{Te}$. The
energy of the initial Maxwellian beam is given by $E_0$. We can then
determine the total energy in the beam and tail electrons at time
$t=100\tau_{ql}$, which indicates the extent to which the beam
relaxation is suppressed and electrons are accelerated. In the
homogeneous case, $E_{beam}=0.76E_0$ and $E_{tail}=0.2E_0$
respectively. In Figure \ref{fig:parst}, we plot the ratios
$E_{beam}/E_{0}$, $E_{tail}/E_{0}$ and $E_{tail}/n_{tail}$ at
$t=100\tau_{ql}$ as functions of the parameter $R$, for a range of
values of $\sqrt{\langle \tilde{n}^2\rangle}$ and ${\mr v}_0$. The
three cases in Figure \ref{fig:w_wo} are marked by asterisks, and
represent three distinct regions in $R$. In the strong inhomogeneity
case, with $\sqrt{\langle \tilde{n}^2\rangle}=1.2\times 10^{-3}$ and
$R\ll1$, the beam-plasma instability is suppressed, so the beam
remains close to its initial Maxwellian form and we find
$E_{beam}=0.99E_0$ and $E_{tail}=0.21E_0$. Very weak inhomogeneity,
with $\sqrt{\langle \tilde{n}^2\rangle}=3.7\times 10^{-4}$
corresponding to $R\gg1$, has little effect on the beam relaxation,
so the beam energy is close to the homogeneous value, but the tail
energy is slightly increased due to the broadened plateau, giving
$E_{beam}=0.78E_0$ and $E_{tail}=0.23E_0$ The intermediate case,
with $\sqrt{\langle \tilde{n}^2\rangle}=3.7\times 10^{-3}$ and
$R\sim 1$, gives $E_{beam}=0.95E_0$ and $E_{tail}=0.35E_0$. Both the
total tail electron energy (middle panel) and the average energy of
a tail electron (right panel), show a significant increase over more
than two orders of magnitude in $R$ around $R\sim 1$. To summarize
we have identified three distinct regimes of the beam plasma
instability in a fluctuating plasma which are controlled by the
parameter $R$. When $R\gg 1$, the density fluctuations have a weak
refractive power and the relaxation proceeds as in a homogeneous
plasma. When $R\ll 1$, diffusive broadening of the wave spectrum is
large, the waves are mainly reabsorbed by the thermal particles,
their level stays small and the instability is suppressed. When
$R\sim 1$ diffusive broadening of the wave spectrum is such that a
substantial part of the wave energy is now also reabsorbed by the
beam electrons leading to acceleration of the beam particles. Using
Eq. (\ref{eqn:diffcoeffGauss}), we may express the condition for
strong acceleration, $R\sim 1$ as
\begin{equation}
\frac{n_{b}}{n_{e}}\left(\frac{\Delta {\mr
v}_{b}}{{\mr v}_{b}}\right)^{2}\sim 
\langle \tilde{n}^{2}\rangle\frac{q_{0}^2 {\mr v}_{b}^2}{{\Omega}_{0}\omega_{pe}}
\left(1+\frac{v_g^2}{v_0^2}\right)^{-3/2}.
\end{equation}

As discussed in Section 2, for a Gaussian spectrum of density
fluctuations there are two extreme regimes of wavenumber diffusion,
where the characteristic group velocity of the beam driven Langmuir
waves, ${\mr v}_g$ is much larger or smaller than ${\mr v}_{0}$. Let
us notice that all of the values for ${\mr v}_0$ used in the above
simulations lie in the transitional region ${\mr v}_g\sim{\mr
v}_{0}$. The effect of ${\mr v}_0$ on the $k$-dependence of the
diffusion coefficient and hence on $\tau_{D}$ is therefore also important, and accounts for the vertical scatter in the points in Figure
\ref{fig:parst}.

\begin{figure}
\includegraphics[width=0.98\linewidth]{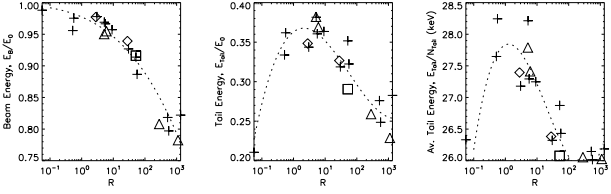}
\caption{As Fig.\ref{fig:parst}, the total energy of beam (left) and tail (middle) electrons, both normalised to the initial beam energy $E_0$, and the average energy of a tail electron (right), all at $t=100\tau_{ql}$ against $R=\tau_D/\tau_{ql}$. Simulation results for  various values of the beam, plasma and density fluctuation parameters are shown. Ranges for $\sqrt{\langle\tilde{n}\rangle}, {\mr v}_0$ are as in the text. For $\omega_{pe}=1$GHz we show: Gaussian fluctuations with $n_b=10^5$ (crosses) and $n_b=10^4$ (squares), powerlaw fluctuations with indices $\frac{5}{3}$ and $\frac{7}{3}$ (triangles). Finally for $\omega_{pe}=200$MHz we show Gaussian fluctuations with $n_b=10^3$ (diamonds). }\label{fig:ParsAll}
\end{figure}

Figure \ref{fig:ParsAll} shows the total beam and tail energies and
average tail electron energy for several values of background
density and beam density, for both Gaussian and power-law density fluctuation spectra. The distribution of the beam and tail total energies with respect to the parameter $R$ is essentially unchanged
from Figure \ref{fig:parst}. We note that collisional effects remain
unimportant for all parameters chosen. Lower beam densities than
those shown produce insufficient levels of Langmuir waves to cause
acceleration, and the parameter $R$ is unaffected by changes in
local plasma frequency. The results for power-law fluctuation
spectra are similar to those for a Gaussian spectrum. We may
therefore expect significant electron acceleration to occur whenever
$R\sim1$, regardless of the specific spectrum of fluctuations, or the specific values of the other parameters.

\section{Generalization to 3D}

Using the same methodology developed for the one-dimensional case,
we now generalize the diffusive description of Langmuir waves to
three-dimensions. The main difference with the 1D case is that in 3D
refraction can change both the absolute value and the orientation of
the wavevector $\vec{k}$. The conservation equation for the
spectral energy density $W(\vec{x},\vec{k},t)$ [erg], is now
\begin{equation} \frac {\partial W(\vec{x}, \vec{k}, t)}{\partial t}
+ \vec{{\mr v}}_g .\nabla W(\vec{x},\vec{k}, t) -\frac{1}{2}
\omega_{pe} \nabla \tilde{n}(\vec{x},t)\cdot\frac{\partial
W(\vec{x}, \vec{k}, t)}{\partial \vec{k}}=0,
\end{equation}
where $\vec{{\mr v}_g}=(3 {\mr v}_{Te}^2/\omega_{pe})\vec{k}$  is the group velocity. In 3D the diffusion equation takes the form
\begin{equation}\label{eq:diff_3d}
\frac{\partial W(\vec{k},t)}{\partial t}=\frac{\partial}{\partial
k_i} D_{ij}   \frac{\partial W(\vec{k},t)}{\partial {k_j}}
\end{equation}
with the diffusion tensor given by
\begin{equation}\label{eqn:genDiff}
D_{ij}(\vec{k})=2\pi \omega_{pe}^2 \int d\Omega\int
\frac{d^3q}{(2\pi)^3}\; q_iq_j S_n(\vec{q}, \Omega)
\delta\left(\Omega-\vec{q}\cdot \vec{{\mr v}_g}\right),
\end{equation}
where $S_n(\vec{q}, \Omega)$ is the spectrum of the density fluctuations and $\int d\mathbf{k} W(\mathbf{k},t)=E$. In the particular case where density fluctuations are due to waves with a dispersion relation $\Omega=\Omega(q)$, then $S_n(\vec{q},\Omega)=S_n(\vec{q})\delta(\Omega-\Omega(\vec{q}))$ and
\begin{equation}\label{eqn:genDiffw}
D_{ij}(\vec{k})=2\pi \omega_{pe}^2\int d^3q\; q_iq_j S_n(\vec{q})
\delta\left(\Omega(\vec{q})-\vec{q}\cdot \vec{{\mr v}_g}\right).
\end{equation}
In spherical coordinates with $\vec{k}=(k,\theta,\phi)$, the general equation for diffusion (assuming azimuthal symmetry) is
\begin{equation}
\frac{\partial W(\vec{k},t)}{\partial
t}=\left(\frac{1}{k^2}\frac{\partial}{\partial
k}\left(k^2D_{kk}\frac{\partial}{\partial k}\right)
+\frac{1}{\sin\theta}\frac{\partial}{\partial \theta}
\sin\theta\left(\frac{D_{\theta\theta}}{k^2}\frac{\partial}{\partial
\theta}\right)\right)W(\vec{k},t)
\end{equation}
with coefficients given in the Appendix below by Equation
\ref{eqn:dSph}.

Expressions for the only non-zero components of the diffusion tensor
$D_{kk}$ and $D_{\theta\theta}$ are also derived in the Appendix
\ref{A1} and are given by
\begin{equation}\label{eqn:Dkkmain} D_{kk}
=\frac{\omega_{pe}^2}{216 \pi {\mr v}_{Te}^3 (k\lambda_{de})^3} \int dq
\Omega(q)^{2}q S_n(q) \end{equation}
\begin{equation}\label{eqn:Dththmain}D_{\theta\theta}=\frac{\omega_{pe}^2}{24
\pi {\mr v}_{Te} k\lambda_{de}}\int dq q^3
\left(1-\left(\frac{\Omega(q)}{3{\mr v}_{Te} k\lambda_{de}
q}\right)^2\right) S_n(q)\end{equation}  Note here that due to the resonance condition $\Omega(q) \le 3 {\mr v}_{Te} k\lambda_{de} q$ the second expression is always positive.

The particular case of elastic scattering is recovered by setting $\Omega=0$ in Equations
(\ref{eqn:Dkkmain}) and (\ref{eqn:Dththmain}) in which case we find that \begin{equation}D_{kk}=0\end{equation} and
\begin{equation}
D_{\theta\theta}=\frac{\omega_{pe}^2}{24 \pi {\mr v}_{Te}
k\lambda_{de}}\int dq q^3 S_n(q),
\end{equation} as given by \citet{1985SoPh...96..181M}. Diffusion then occurs in angle only at a rate independent of
$\theta$ and tends to isotropise the Langmuir wave spectrum. 

It is
instructive to rewrite the diffusion equation in terms of the
independent variables $k$ and $k_\parallel$, where $k_\parallel$ is
the component of the wavevector parallel to the beam direction,
given by $k_\parallel=k\cos\theta$. In the limit $\cos\theta\sim 1$,
or equivalently $k\sim k_\parallel$, we obtain
\begin{equation} \label{eqn:diffkpar} \frac{\partial}{\partial t}
W(k_\parallel,t)=\frac{1}{k_\parallel^2}\frac{\partial}{\partial
k_\parallel}\left(k_\parallel^2D_{kk}\frac{\partial}{\partial
k_\parallel}\right)W(k_\parallel,t)
-2\frac{D_{\theta\theta}}{k_\parallel}  \frac{\partial}{\partial
k_\parallel} W(k_\parallel,t).
\end{equation}
This shows that both magnitude and angular change in wave-vector, as
given by $D_{kk}$ and $D_{\theta\theta}$, contribute to the
redistribution of wave energy in wavenumber \emph{parallel} to the
beam. This transport of wave energy is diffusive for the former and
convective for the latter. In both cases, waves shifted to smaller
parallel wavenumbers can be absorbed by beam electrons at larger
parallel velocities resulting in acceleration of tail electrons. A
large body of work has been devoted to the role of angular diffusion
\citep[e.g.][]{1976JPSJ...41.1757NF} with respect to the suppression
of the beam-plasma instability, mainly in the context of elastic
scattering, i.e. $D_{kk}=0$. The electron distribution was assumed to be fixed and the effect of
scattering of waves by density fluctuations was modeled solely as
wave damping. However, the energy lost by the waves due to damping
is transferred to the electrons leading to electron acceleration. 

A full numerical treatment of this 3D diffusion and its acceleration effects is not possible using our 1D quasilinear code, and projecting the 3D case into 1D is uninformative. However, as seen in the previous section, acceleration of beam
electrons in the 1D case occurs when the transport time scale in $k$-space is of the
order of the the quasilinear time scale, for a wide range of beam, plasma and fluctuation parameters. Angular diffusion
convects wave energy in $k_{\parallel}$, always toward small
$k_{\parallel}$ and on a time scale given by
$\tau_{\theta\theta}=k_0^2/D_{\theta\theta}$ (see Equation
(\ref{eqn:diffkpar})). Thus we expect angular diffusion to also lead to
electron acceleration, confirmed by, for example, the PIC simulations of \citet{2012A&A...544A.148K}, which include the effect of long wavelength density fluctuations on Langmuir waves, and find such an acceleration effect. From our diffusion treatment we may estimate that this effect will be significant when $\tau_{\theta\theta}\sim \tau_{ql}$,
whether scattering is elastic or not. It is also natural to
expect that a necessary condition for the electron distribution
function to remain fixed, and hence the beam-plasma relaxation to be
suppressed, is that $\tau_{\theta\theta}\ll \tau_{ql}$ otherwise
when $\tau_{\theta\theta}\sim \tau_{ql}$ the effect of electron
acceleration studied in this work needs also to be considered in the
overall wave-particle energy budget. 


\section{Summary}
In summary, we have considered the effects of the diffusion of
Langmuir waves in wavenumber space numerically, using a
1-dimensional model. We found the potential to suppress the
beam-plasma instability when the diffusion is sufficiently fast, but
also the possibility of a significant acceleration of beam
electrons. The transition from acceleration to suppression is
controlled by the ratio of the quasilinear and
$k$-space diffusive timescales, with the most efficient acceleration occurring
when the quasilinear time is close to the timescale for diffusion of Langmuir waves in wavenumber space. 
This relationship was found to hold for a wide range of beam, plasma, and fluctuation parameters. 
In addition, while not presented here, previous work found a similar acceleration effect to
occur during the collisional relaxation of a power law beam
\citep{2012A&A...539A..43K}. We have also derived the
diffusion coefficients for the general case of inelastic scattering
in 3-dimensions. This constitutes an extension of the elastic
scattering theory previously discussed in the literature. A
coefficient $D_{kk}\propto \Omega^2$ and a correction to the
coefficient $D_{\theta\theta}$ proportional to $\Omega^2$ were
found. From the diffusion equation and the effects of diffusion in
1D, we infer that scattering in 3D will be capable of producing
electron acceleration, provided the transport time-scale in
wave-vector space is of the order of the quasilinear timescale,
while the beam plasma relaxation is suppressed only under the
condition that the former is much smaller than the later.

\appendix
\section{Cartesian and Spherical representations of the diffusion tensor}\label{A1}

The diffusion equation in Cartesian coordinates is
\begin{equation}\label{eq:diff_3d_app}
\partial_t W(\vec{k},t)=\partial_{k_i} D_{ij} \partial_{k_j} W(\vec{k},t)
\end{equation}
with coefficient given by
\begin{equation}\label{eqn:genDiff_app}
D_{ij}(\vec{k})=2\pi \omega_{pe}^2 \int d\Omega\int \frac{d^3q}{(2\pi)^3}\; q_iq_j S_n(\vec{q}, \Omega) \delta\left(\Omega-\vec{q}\cdot \vec{v_g}\right)
\end{equation}
where $S_n(\vec{q}, \Omega)$ is the spectrum of the density fluctuations.

\subsection{Definition of coordinates}
In Cartesian coordinates, we define one axis to be parallel to the beam direction, labelled as $\parallel$, and have two mutually perpendicular axes labelled, $\perp_1, \perp_2$, giving a standard right-handed Cartesian coordinate system. We also define a spherical coordinate system, with $\theta$ the angle to the beam direction, and $\phi$ the azimuth, measured clockwise around the beam direction.

In these spherical coordinates we write the Langmuir wavevector as $\vec{k}=(k, \theta, \phi)$. Assuming azimuthal symmetry, we may define the coordinates such that the azimuth of $\vec{k}$ is zero. The diffusion tensor can then be transformed from the Cartesian coordinates, in which it has components $D_{ij}$ ($D_{\parallel\parallel}, D_{\parallel\perp_1}, D_{\parallel\perp_2}$ etc) as given by Equation \ref{eqn:genDiff_app}, into these spherical coordinates.
The diffusion equation may also be transformed, becoming
\begin{equation}
\partial_{k_i} D_{ij} \partial_{k_j} W(\vec{k},t)=\left(\frac{1}{k^2}\partial_k\left(k^2D_{kk}\partial_k+kD_{k\theta}\partial_\theta\right) +\frac{1}{\sin\theta}\partial_\theta \sin\theta\left(\frac{D_{\theta\theta}}{k^2}\partial_\theta+\frac{D_{\theta k}}{k}\partial_k\right) \right)W(\vec{k},t)
\end{equation} There is no diffusion in $\phi$ due to azimuthal symmetry.

By analogy with the Cartesian expression in Equation \ref{eqn:genDiff_app}, we define the following quantities

\begin{equation}\label{eqn:dthetatheta}
q_{\theta\theta}= q_{\perp_1}q_{\perp_1}\cos^2\theta- 2q_{\perp_1}q_\parallel\sin\theta\cos\theta +q_{\parallel}q_{\parallel}\sin^2\theta
\end{equation}
\begin{equation}\label{eqn:dkk}
q_{kk}=q_{\perp_1}q_{\perp_1}\sin^2\theta + 2q_{\perp_1}q_{\parallel}\sin\theta\cos\theta +q_{\parallel}q_{\parallel}\cos^2\theta
\end{equation}
and
\begin{equation}\label{eqn:dktheta}
q_{k\theta}=q_{\theta k}=\sin\theta\cos\theta (q_{\perp_1}q_{\perp_1}-q_{\parallel}q_{\parallel})+(\cos^2\theta-\sin^2\theta)q_{\perp_1}q_{\parallel}
\end{equation} which allow us to write simply

\begin{equation}\label{eqn:dSph_cart}
D_{ij}(\vec{k})=2\pi \omega_{pe}^2 \int d\Omega\int \frac{q^2}{(2\pi)^3}dq\int d\bar{\mu}\int d\bar{\phi}\; q_{ij} S_n(\vec{q}, \Omega) \delta\left(\Omega-\vec{q}\cdot \vec{v}_g\right)
\end{equation}
where $i,j =\theta, k, \phi$.

\subsection{Expressions for the diffusion coefficients}

Assuming a dispersion relation $\Omega=\Omega(q)$ (the general case being intractable) we write $S_n(\vec{q},\Omega)=S_n(\vec{q})\delta(\Omega-\Omega(q))$ so that Equation \ref{eqn:dSph_cart} becomes
\begin{equation}\label{eqn:dSph}
D_{ij}(\vec{k})=2\pi \omega_{pe}^2 \int\frac{q^2}{(2\pi)^3}dq\int d\bar{\mu}\int d\bar{\phi}\; q_{ij} S(\vec{q}) \delta\left(\Omega-\vec{q}\cdot \vec{v}_g\right)
\end{equation}
Substituting
\begin{equation}\vec{k}\cdot\vec{q}=kq \left((1-\mu^2)^{1/2}(1-\bar{\mu}^2)^{1/2}\cos\bar{\phi}+\mu\bar{\mu}\right)
\end{equation} into the delta function, and integrating over $\bar{\phi}$, the resonance condition  becomes \begin{equation}
 \cos \bar{\phi}=\frac{\Omega'-\mu\bar{\mu}}{(1-\mu^2)^{1/2}(1-\bar{\mu}^2)^{1/2}}
 \end{equation} where $\Omega'(q)=\Omega(q)\omega_{pe}/(3v_{Te}^2 k q)$ (resonance also requires $\Omega'\le 1$).

We use this to evaluate Equations \ref{eqn:dthetatheta} to \ref{eqn:dktheta}, finding \[q_{k\theta}=0\] \[q_{kk}=q^2\Omega'(q)^2\] \[q_{\theta\theta}=q^2 \left(\bar{\mu}^2+\mu^2\Omega'(q)^2-2\mu\bar{\mu}\Omega'(q)\right)/(1-\mu^2)\]
The diffusion coefficients are then \begin{equation}
D_{kk}=\frac{\omega_{pe}^3}{12 \pi^2 v_{Te}^2 k}  \int dq q^3 \int d\bar{\mu}\Omega'(q)^2 S_n(\vec{q}) \left( (1-\mu^2)(1-\bar{\mu}^2) -\Omega^{'2} + 2\mu\bar{\mu}\Omega' -\mu^2\bar{\mu}^2\right)^{-1/2}
\end{equation} and
\begin{equation}D_{\theta\theta}=\frac{\omega_{pe}^3}{12 \pi^2 v_{Te}^2 k}  \int dq q^3 \int d\bar{\mu}\frac{\bar{\mu}^2+\mu^2\Omega'(q)^2-2\mu\bar{\mu}\Omega'(q)}{1-\mu^2} S_n(\vec{q}) \left( (1-\mu^2)(1-\bar{\mu}^2) -\Omega^{'2} + 2\mu\bar{\mu}\Omega' -\mu^2\bar{\mu}^2\right)^{-1/2}
\end{equation}
where the limits on the $\bar{\mu}$ integrals are the solutions of $(1-\mu^2)(1-\bar{\mu}^2) -\Omega^{'2} + 2\mu\bar{\mu}\Omega' -\mu^2\bar{\mu}^2=0$ i.e $\mu_\pm=\mu\Omega' \pm (1-\mu^2)^{1/2}(1-\Omega^{'2})^{1/2}$ (noting that $\Omega'\le 1$).

Assuming isotropic fluctuations, the $\bar{\mu}$ integrals can be calculated, and the results are
\begin{equation}\label{eqn:diffkk}D_{kk} =\frac{\omega_{pe}^2}{216 \pi v_{Te}^3 (k\lambda_{de})^3} \int dq q\Omega(q)^{2} S_n(q)  \end{equation}  \begin{equation}\label{eqn:diffthetatheta}D_{\theta\theta}=\frac{\omega_{pe}^2}{24 \pi v_{Te} k\lambda_{De}}\int dq q^3 \left(1-\left(\frac{\Omega(q)}{3v_{Te} k\lambda_{De} q}\right)^2\right) S_n(q).\end{equation}

The special case of elastic scattering, that was treated by \citet{1976JPSJ...41.1757NF, 1985SoPh...96..181M}
 may be simply recovered by taking $\Omega=0$, for which we find \begin{equation}D_{kk}=0\end{equation} \begin{equation}D_{\theta\theta}=\frac{\omega_{pe}^3}{24 \pi v_{Te}^2 k}\int dq q^3 S_n(q) \end{equation}


\begin{thebibliography}{37}
\bibitem[{{Berezin} {et~al.}(1964){Berezin}, {Berezina}, {Bolotin}, \&
  {Fainberg}}]{1964JNuE....6..173B}
{Berezin}, A.~K., {Berezina}, G.~P., {Bolotin}, L.~I., \& {Fainberg}, Y.~B.
  1964, Journal of Nuclear Energy, 6, 173

\bibitem[{{Bian} \& {Kontar}(2010)}]{2010PhPl...17f2308B}
{Bian}, N.~H. \& {Kontar}, E.~P. 2010, Physics of Plasmas, 17, 062308

\bibitem[{{Bian} {et~al.}(2010){Bian}, {Kontar}, \&
  {Brown}}]{2010A&A...519A.114B}
{Bian}, N.~H., {Kontar}, E.~P., \& {Brown}, J.~C. 2010, \aap, 519, A114

\bibitem[{{Bre{\v i}zman} \& {Ryutov}(1969)}]{1969JETP...30..759B}
{Bre{\v i}zman}, B.~N. \& {Ryutov}, D.~D. 1969, Soviet Journal of Experimental
  and Theoretical Physics, 30, 759

\bibitem[{{Celnikier} {et~al.}(1983){Celnikier}, {Harvey}, {Jegou}, {Moricet},
  \& {Kemp}}]{1983A&A...126..293C}
{Celnikier}, L.~M., {Harvey}, C.~C., {Jegou}, R., {Moricet}, P., \& {Kemp}, M.
  1983, \aap, 126, 293

\bibitem[{{Cronyn}(1972)}]{1972ApJ...171L.101C}
{Cronyn}, W.~M. 1972, \apjl, 171, L101

\bibitem[{{Escande}(1975)}]{1975PhRvL..35..995E}
{Escande}, D.~F. 1975, Physical Review Letters, 35, 995

\bibitem[{{Escande} \& {de Genouillac}(1978)}]{1978A&A....68..405E}
{Escande}, D.~F. \& {de Genouillac}, G.~V. 1978, \aap, 68, 405

\bibitem[{{Ginzburg} \& {Zhelezniakov}(1958)}]{1958SvA.....2..653G}
{Ginzburg}, V.~L. \& {Zhelezniakov}, V.~V. 1958, Soviet~Ast., 2, 653

\bibitem[{{Goldman} \& {Dubois}(1982)}]{1982PhFl...25.1062G}
{Goldman}, M.~V. \& {Dubois}, D.~F. 1982, Physics of Fluids, 25, 1062

\bibitem[{{Hannah} {et~al.}(2009){Hannah}, {Kontar}, \&
  {Sirenko}}]{2009ApJ...707L..45H}
{Hannah}, I.~G., {Kontar}, E.~P., \& {Sirenko}, O.~K. 2009, \apjl, 707, L45

\bibitem[{{Karlick{\'y}} \& {Kontar}(2012) {Karlick{\'y}} \& {Kontar}}]{2012A&A...544A.148K}
{Karlick{\'y}}, M. and {Kontar}, E.~P., \aap, 544, A148

\bibitem[{{Kontar}(2001)}]{2001A&A...375..629K}
{Kontar}, E.~P. 2001, \aap, 375, 629

\bibitem[{{Kontar} \& {P{\'e}cseli}(2002)}]{2002PhRvE..65f6408K}
{Kontar}, E.~P. \& {P{\'e}cseli}, H.~L. 2002, \pre, 65, 066408

\bibitem[{{Kontar} {et~al.}(2012){Kontar}, {Ratcliffe}, \&
  {Bian}}]{2012A&A...539A..43K}
{Kontar}, E.~P., {Ratcliffe}, H., \& {Bian}, N.~H. 2012, \aap, 539, A43

\bibitem[{{Krasovskii}(1978)}]{1978SvJPP...4R1267K}
{Krasovskii}, V.~L. 1978, Soviet Journal of Plasma Physics, 4, 1267

\bibitem[{{Lifshitz} \& {Pitaevskii}(1981)}]{1981phki.book.....L}
{Lifshitz}, E.~M. \& {Pitaevskii}, L.~P. 1981, {Physical kinetics}, ed.
  {Lifshitz, E.~M.~\& Pitaevskii, L.~P.}

\bibitem[{{Liperovskii} \& {Tsytovich}(1965)}]{1965JAMTP...6....9L}
{Liperovskii}, V.~A. \& {Tsytovich}, V.~N. 1965, Journal of Applied Mechanics
  and Technical Physics, 6, 9

\bibitem[{{Melrose}(1974)}]{1974SoPh...35..441M}
{Melrose}, D.~B. 1974, Sol.~Phys., 35, 441

\bibitem[{{Melrose}(1987)}]{1987SoPh..111...89M}
{Melrose}, D.~B. 1987, Sol.~Phys., 111, 89

\bibitem[{{Melrose}(1990)}]{1990SoPh..130....3M}
{Melrose}, D.~B. 1990, Sol.~Phys., 130, 3

\bibitem[{{Muschietti} \& {Dum}(1991)}]{1991PhFlB...3.1968M}
{Muschietti}, L. \& {Dum}, C.~T. 1991, Physics of Fluids B, 3, 1968

\bibitem[{{Muschietti} {et~al.}(1985){Muschietti}, {Goldman}, \&
  {Newman}}]{1985SoPh...96..181M}
{Muschietti}, L., {Goldman}, M.~V., \& {Newman}, D. 1985, \solphys, 96, 181

\bibitem[{{Nishikawa} \& {Ryutov}(1976)}]{1976JPSJ...41.1757NF}
{Nishikawa}, K. \& {Ryutov}, D.~D. 1976, Journal of the Physical Society of
  Japan, 41, 1757

\bibitem[{{Papadopoulos}(1975)}]{1975PhFl...18.1769P}
{Papadopoulos}, K. 1975, Physics of Fluids, 18, 1769

\bibitem[{{Reid} \& {Kontar}(2010)}]{2010ApJ...721..864R}
{Reid}, H.~A.~S. \& {Kontar}, E.~P. 2010, ApJ, 721, 864

\bibitem[{{Reid} {et~al.}(2011){Reid}, {Vilmer}, \&
  {Kontar}}]{2011A&A...529A..66R}
{Reid}, H.~A.~S., {Vilmer}, N., \& {Kontar}, E.~P. 2011, \aap, 529, A66+

\bibitem[{{Robinson}(1983)}]{1983PASAu...5..208R}
{Robinson}, R.~D. 1983, Proceedings of the Astronomical Society of Australia,
  5, 208

\bibitem[{{Sturrock}(1966)}]{1966PhRv..141..186S}
{Sturrock}, P.~A. 1966, Physical Review, 141, 186

\bibitem[{{Tsiklauri}(2010)}]{2010SoPh..267..393T}
{Tsiklauri}, D. 2010, \solphys, 267, 393

\bibitem[{{Tsytovich}(1995)}]{1995lnlp.book.....T}
{Tsytovich}, V.~N. 1995, {Lectures on Non-linear Plasma Kinetics}, ed.
  {Tsytovich, V.~N.~\& ter Haar, D.}

\bibitem[{{Vedenov} {et~al.}(1967){Vedenov}, {Gordeev}, \&
  {Rudakov}}]{1967PlPh....9..719V}
{Vedenov}, A.~A., {Gordeev}, A.~V., \& {Rudakov}, L.~I. 1967, Plasma Physics,
  9, 719

\bibitem[{{Vedenov} \& {Velikhov}(1963)}]{1963JETP...16..682V}
{Vedenov}, A.~A. \& {Velikhov}, E.~P. 1963, Soviet Journal of Experimental and
  Theoretical Physics, 16, 682

\bibitem[{{Whitham}(1965)}]{1965JFM....22..273W}
{Whitham}, G.~B. 1965, Journal of Fluid Mechanics, 22, 273

\bibitem[{{Yoon} {et~al.}(2006){Yoon}, {Rhee}, \& {Ryu}}]{2006JGRA..11109106Y}
{Yoon}, P.~H., {Rhee}, T., \& {Ryu}, C.-M. 2006, Journal of Geophysical
  Research (Space Physics), 111, 9106

\bibitem[{{Zakharov}(1974)}]{1974R&QE...17..326Z}
{Zakharov}, V.~E. 1974, Radiophysics and Quantum Electronics, 17, 326

\bibitem[{{Zheleznyakov} \& {Zaitsev}(1970)}]{1970SvA....14...47Z}
{Zheleznyakov}, V.~V. \& {Zaitsev}, V.~V. 1970, \sovast, 14, 47

\bibitem[{{Ziebell} {et~al.}(2011){Ziebell}, {Yoon}, {Pavan}, \&
  {Gaelzer}}]{2011ApJ...727...16Z}
{Ziebell}, L.~F., {Yoon}, P.~H., {Pavan}, J., \& {Gaelzer}, R. 2011, \apj, 727,
  16

\end{thebibliography}

\end{document}